\documentclass[prl,twocolumn,floatfix,showpacs ]{revtex4-1}
\UseRawInputEncoding
\usepackage{amsmath}
\usepackage{amssymb}
\usepackage{graphicx}
\usepackage{dcolumn}
\usepackage{bm}
\usepackage[colorlinks=true,linkcolor=blue,anchorcolor=blue, citecolor=cyan,urlcolor=cyan]{hyperref}
\usepackage[mathlines]{lineno}
\usepackage{ulem}
\usepackage{epstopdf}
\begin{document}
\title{Proposal of a general scheme: valley  polarization  in  antiferromagnetic bilayer systems}
\author{San-Dong Guo$^{1}$}
\email{sandongyuwang@163.com}
\author{Ping Li$^{2}$ and  Guangzhao Wang$^{3}$}
\affiliation{$^{1}$School of Electronic Engineering, Xi'an University of Posts and Telecommunications, Xi'an 710121, China}
\affiliation{$^2$State Key Laboratory for Mechanical Behavior of Materials, School of Materials Science and Engineering, Xi'an Jiaotong University, Xi'an, Shaanxi 710049, China}
\affiliation{$^3$Key Laboratory of Extraordinary Bond Engineering and Advanced Materials Technology of Chongqing, School of Electronic Information Engineering, Yangtze Normal University, Chongqing 408100, China}
\begin{abstract}
Superior to ferromagnetic (FM) valleytronics, antiferromagnetic (AFM) counterpart
 exhibits  ultradense and ultrafast potential due to their intrinsic advantages of zero stray field, terahertz dynamics, and compensated moment of  antiferromagnets.
However, the physics of  spontaneous valley polarization  is mainly rooted in FM hexagonal lattices and is rarely used to
explore the simultaneous spin and valley polarizations in AFM materials.
Here, we propose a general stacking way to achieve valley  polarization  in  AFM bilayer systems.
The hexagonal ferrovalley material is used as the basic building unit, and then the space-inversion centrosymmetric bilayer system with interlayer AFM ordering is constructed by horizontal mirror and 2-fold rotational operations, which can exhibit spontaneous valley polarization. In this construction process, the rarely explored \textit{layer-locked hidden valley polarization}, hidden Berry curvature and layer Hall effect are involved, and  an out-of-plane electric field can be used to detect hidden valley polarization and to realize layer-locked anomalous valley Hall effect. We use three examples to illustrate our proposal. Firstly, the Janus GdBrI is used to prove  concepts and effects involved in our design process. Secondly, the $\mathrm{RuBr_2}$ is used to demonstrate other phenomena, including valley polarization transition and \textit{near-ideal quantum spin Hall insulator}. Finally, we use our design principles to understand the valley polarization of experimentally synthesized MnSe from a new perspective. Our
works establish a robust general scheme to achieve valley  polarization  in  AFM bilayer systems, thereby
opening up new avenues for AFM valleytronics.

\end{abstract}
\maketitle
\textcolor[rgb]{0.00,0.00,1.00}{\textbf{Introduction.---}}
In addition to charge and spin, the valley
degree of freedom in valleytronic materials can be used to  process information and to perform  logic operations with the advantage of low power consumption and high speed\cite{q1,q2,q4}.
Due to the large separation in momentum space, valley, characterizing the energy extrema of band, is robust against smooth deformations and low-energy phonons. The current focus of valley research is how to stably manipulate carriers and break the balance  in the valleys, thereby giving rise to robust valley polarization\cite{v5,v9,v10,v11}.
In general, the valley physics is concentrated at -K/K point in the  Brillouin zone (BZ) for two-dimensional (2D) hexagonal lattice,
and time inversion symmetry ($T$) in nonmagnetic materials leads to the disappearance of valley polarization: $E_{\uparrow}(K)$=$T$$E_{\uparrow}(K)$= $E_{\downarrow}(-K)$. The 2D magnetic materials with broken $T$ symmetry can serve as an ideal platform for facilitating valley polarization, and the concept of ferrovalley
 with spontaneous valley polarization has
been firstly proposed, which can  achieve anomalous valley Hall  effect (AVHE)\cite{q10}.

In the past few years, the research on ferrovalley mainly focuses on 2D ferromagnetic (FM) hexagonal lattice to explore spontaneous valley polarization\cite{q11,q12,q13,q13-1,q14,q14-1,q15,q16,q18}.
 Compared with ferromagnets, antiferromagnetic (AFM) candidates have more advantages in valleytronic applications  due to  the high storage density, robustness against external magnetic field, as well as the ultrafast writing speed\cite{k1,k2}. Thus, it
is of fundamental importance and high interest to seek out 2D AFM materials with valley polarization, accompanied by AVHE.
Collinear antiferromagnets can be divided into three categories by critical symmetries: inversion symmetry ($P$), translational symmetry ($\tau$) and other symmetries (for example: rotational or mirror  symmetry).
The first two classes are spin degenerate, while the third class possesses spin splitting,  called altermagnetism\cite{k4,k5}.
This spontaneous valley polarization accompanied by AVHE has been explored in the first class of antiferromagnets with combined symmetry ($PT$) of $P$ and $T$\cite{gsd1,gsd2}.
For tetragonal altermagnets, the valley polarization can be achieved by simply breaking the corresponding rotational lattice
symmetry with tuning magnetization direction or apllying uniaxial strain\cite{k60,k6,k7,k7-1,k7-2,k7-3}.
For twisted altermagnetism\cite{k8,k80}, an out-of-plane external electric field can be used to induce valley polarization due to valley-layer coupling\cite{k9,k10}.

\begin{figure*}[t]
    \centering
    \includegraphics[width=0.80\textwidth]{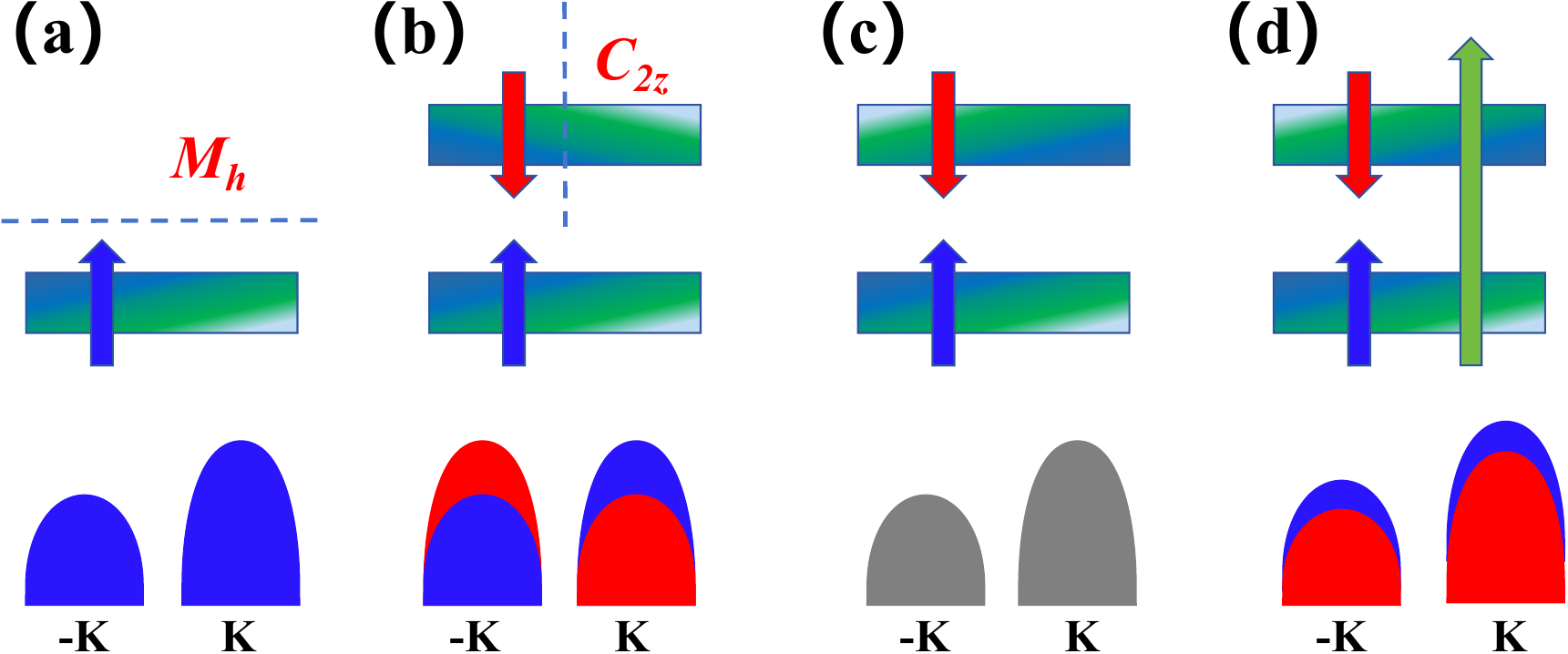}
    \caption{(Color online) Illustration of a general scheme to produce  valley  polarization  in  AFM bilayer systems. (a): A hexagonally symmetrical FM  monolayer with valley  polarization  is employed as the fundamental building block, defined as lower layer. (b): The upper layer can be derived from the lower layer through a mirror operation with respect to the horizontal dashed line in (a), forming a bilayer structure with horizontal mirror symmetry $M_h$. Assuming an AFM ordering, the bilayer system shows no valley  polarization, but possesses hidden valley polarization. (c): The upper  layer is rotated by 180$^{\circ}$ along vertical 2-fold axis in (b), and then the bilayer structure possesses inversion symmetry $P$, giving rise to valley polarization with spin or layer degeneracy. (d): When an out-of-plane electric field is applied, this spin or layer degeneracy is removed, inducing layer-locked valley  polarization. In (a, b, c, d), the blue (red) arrows represent spin-up (spin-down), and the blue (red) valley areas  show characteristic projection of lower layer/spin-up (upper layer/spin-down). In (d), the green arrow represents an out-of-plane electric field. }\label{sy}
\end{figure*}

The twisted altermagnetism provides an
experimentally feasible approach to achieve altermagnetism
 by Van der Waals (vdW) stacking of 2D materials, which takes one of all five 2D Bravais lattices\cite{k8}.
 A natural question is whether vdW stacking of 2D materials can realize  an ideal platform for AFM valley polarization.
 In this work, we propose a general stacking way  for the emergence of $PT$-AFM valley polarization in bilayer
systems by applying symmetry operations based on ferrovalley monolayer.
 Through first-principles calculations, we present
three representative examples demonstrating the feasibility of our proposal.
During this validation, some interesting concepts or effects are involved, such as the \textit{hidden valley polarization}, hidden Berry curvature, layer Hall effect and \textit{near-ideal quantum spin Hall insulator} (QSHI). From our proposed design principles, the valley polarization of  experimentally synthesized MnSe can be intuitively understood.
 Therefore, the AFM valley polarization by vdW stacking  would
bring rich designability to valleytronics.

\textcolor[rgb]{0.00,0.00,1.00}{\textbf{Approach to valley  polarization  in  antiferromagnetic bilayer systems.---}}
The spontaneous valley polarization has been predicted  in a significant array of 2D hexagonal FM materials with  valley physics at -K/K  high symmetry point\cite{q10,q11,q12,q13,q13-1,q14,q14-1,q15,q16,q18}. Here, we employ hexagonally symmetric ferrovalley material as the fundamental building block to elucidate our design principles. In fact, our design principles can be applied to  FM valley-polarized materials with other lattice symmetry as building unit. As shown in \autoref{sy}(a), the ferrovalley monolayer is defined as lower layer, giving rise to valley polarization.

 Initially, the approach
 starts with stacking into a bilayer structure with interlayer AFM ordering (\autoref{sy}(b)), and the upper layer can be derived from the lower layer through a mirror operation with respect to the horizontal dashed line.  The 2D BZ of the lower and upper layers are related by horizontal mirror $M_h$ symmetry, and their $k$ points correspond to each other ($K_{upper}$=$K_{lower}$ and $-K_{upper}$=$-K_{lower}$). Since the two layers have opposite magnetization (layer-spin locking), their valley polarizations are also opposite: the lower layer is K-valley polarized, while the upper layer is -K-valley polarized. These result in that the  bilayer system possesses the rarely explored \textit{layer-locked hidden valley polarization}, along with layer-locked spin polarization. However, no net valley polarization can be observed, which can be explained by symmetry analysis. When consideing spin-orbital coupling (SOC),  the  bilayer system has magnetic group symmetry $M_hT$, which leads to: $E_{\uparrow}(K)$=$M_hT$$E_{\uparrow}(K)$= $M_h$$E_{\downarrow}(-K)$=$E_{\downarrow}(-K)$. Experimentally, the hidden valley polarization can be observed through an out-of-plane electric field, which can introduce the layer-dependent electrostatic potential, staggering the energy bands of the two layers.
\begin{figure*}[t]
    \centering
    \includegraphics[width=0.96\textwidth]{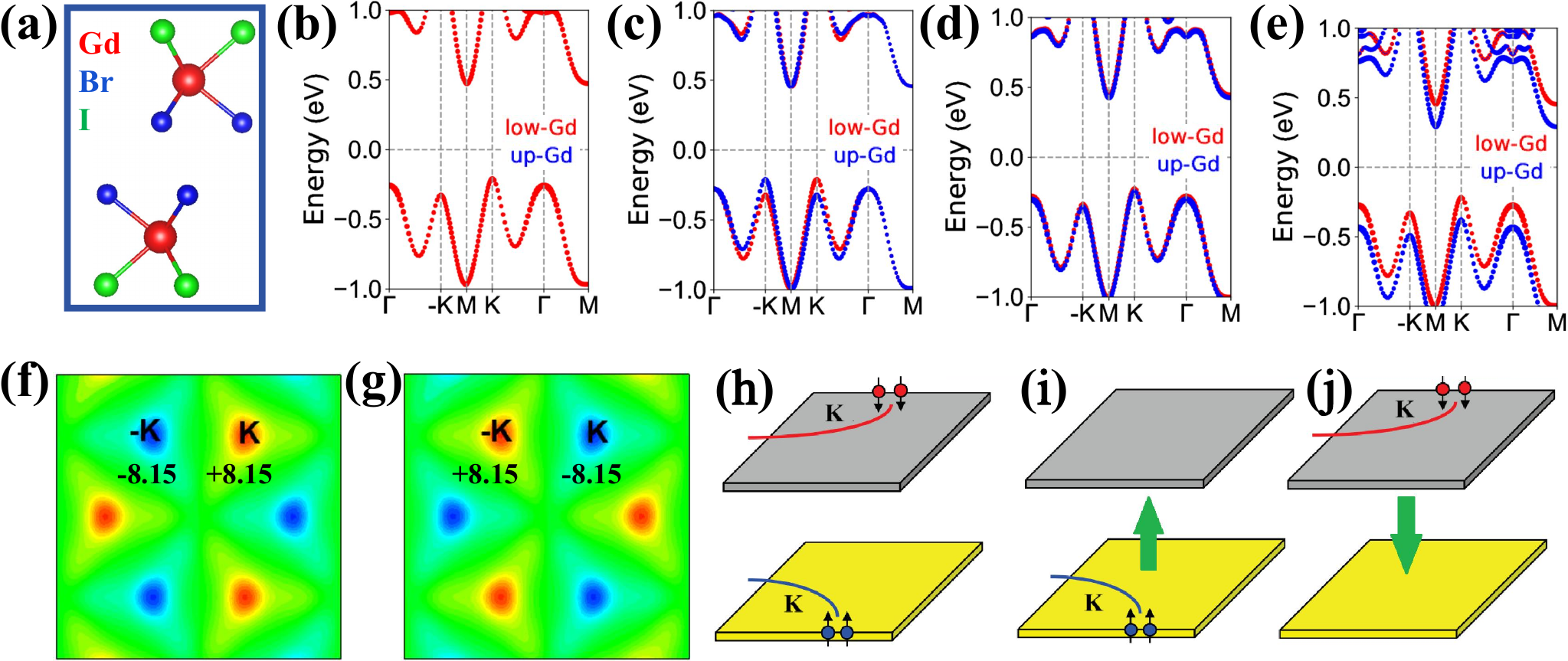}
    \caption{(Color online) (a): The crystal structure of  bilayer Janus GdBrI with $P$ symmetry. (b, c, d, e): The layer-Gd-characteristics  projected  energy band structures of GdBrI for  monolayer,  $M_h$-symmetrical bilayer,  $P$-symmetrical bilayer  and   $P$-symmetrical bilayer  with electric field amplitude of 0.02  $\mathrm{V/{\AA}}$ in the $+z$ direction within SOC. (f, g): The Berry curvature distribution of $P$-symmetrical bilayer GdBrI for the spin-up and spin-down channels. (h, i, j): The valley layer-spin hall effect, and layer-locked anomalous valley Hall effect with $+z$ and $-z$ electric field. In  (b, c, d, e), the red (blue) shows the layer-Gd-characteristics projection of lower layer (upper layer). In (i, j), the green arrows represent the $+z$ and $-z$  electric field.  }\label{gd}
\end{figure*}

 Subsequently, the upper layer is rotated by 180$^{\circ}$ along vertical 2-fold axis (rotation
operation $C_{2z}$), and then the bilayer structure possesses inversion symmetry $P$=$C_{2z}$$M_h$ (\autoref{sy}(c)).
The 2D  BZ of the lower and upper layers are connected by $P$ symmetry, producing that $K_{upper}$=$-K_{lower}$ and $-K_{upper}$=$K_{lower}$. Together with the oppose valley polarization of  two layers, the bands of the two layers coincide exactly with each other,  giving rise to valley polarization with spin or layer degeneracy. This  magnetic group symmetry $PT$ only ensures spin degeneracy: $E_{\uparrow}(K)$=$PT$$E_{\uparrow}(K)$= $P$$E_{\downarrow}(-K)$=$E_{\downarrow}(K)$, and does not connect the -K and K valleys, allowing valley polarization.

Through the above process, the valley  polarization  in AFM bilayer systems can be achieved. Our approach can  apply to all interlayer AFM materials with  the fundamental building block taking any 2D Bravais lattice. Instead of  vdW heterostructure approach,  two identical monolayers can also interact through other strong covalent bonds or intercalations.
An out-of-plane electric field can be used to break $PT$  symmetry, and  the degeneracy of  spin-up and
spin-down bands is removed at each $k$ point, giving rise to spin splitting of $s$-wave symmetry ((\autoref{sy}(d))).
The order of spin splitting can be switched by the reversal of
the direction of  electric field. Strictly speaking, the bilayer system with an applied electric field is not an antiferromagnet, but a fully-compensated ferrimagnet with  net zero magnetic moment, which can  exhibit  magneto-optical effect,  anomalous Hall effect  and fully spin-polarized currents.

\textcolor[rgb]{0.00,0.00,1.00}{\textbf{Material realization.---}}Here, we employ three fundamental examples to exemplify our proposal: (1) Janus GdBrI monolayer as building block, (2)  $\mathrm{RuBr_2}$ monolayer as building block, and (3) experimentally synthesized MnSe. Within density functional theory (DFT)\cite{1},  we perform the spin-polarized  first-principles calculations  within Vienna abinitio Simulation Package (VASP)\cite{pv1,pv2,pv3} by using the projector augmented-wave (PAW) method. The exchange-correlation potential is adopted by generalized gradient
approximation  of Perdew-Burke-Ernzerhof (PBE-GGA)\cite{pbe}.
To account for electron correlation of 3$d$ of $4f$ orbitals, a Hubbard correction is employed within the
rotationally invariant approach proposed by Dudarev et al\cite{du}. The dispersion-corrected DFT-D3 method\cite{dft3} is adopted to describe the vdW
interactions. The Berry curvatures
are calculated directly from the calculated
wave functions  based on Fukui's
method\cite{bm} by using VASPBERRY code\cite{bm3}. The  topological properties  are calculated by using the maximal localized
Wannier function tight-binding model by WANNIER90
and WANNIERTOOLS\cite{w1,w2}.
\begin{figure*}[t]
    \centering
    \includegraphics[width=0.96\textwidth]{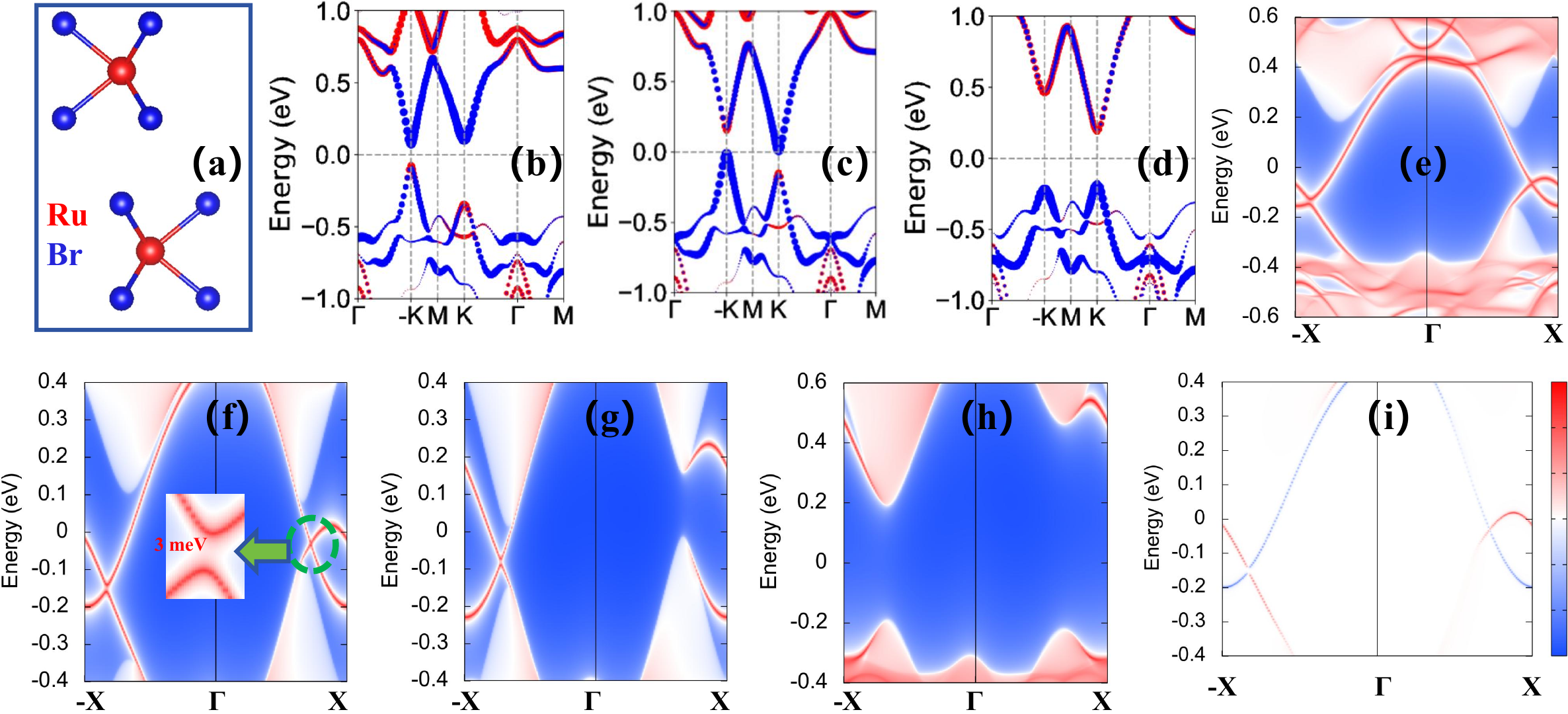}
     \caption{(Color online)For bilayer  $\mathrm{RuBr_2}$ with $P$ symmetry, (a): The crystal structure.  (b, c, d): The Ru-$d_{x^2-y^2}$/$d_{xy}$+$d_{z^2}$-orbital characters of energy bands  with  $U=$1.5 eV, 2.0 eV and 2.5 eV by including SOC. (e, f, g, h): The edge states  calculated  along the (100) direction with $U=$1.0 eV, 1.5 eV, 2.0 eV and 2.5 eV. (i): The spin components of the edge states with $U=$1.5 eV. In (b, c, d), the red (blue) represents  $d_{x^2-y^2}$/$d_{xy}$ ($d_{z^2}$)-orbital characters. In (i),  the red (blue) represents spin-up (spin-down) characters.}\label{ru}
\end{figure*}

(1) Janus GdBrI monolayer: The Janus 2H-GdXY (X, Y=Cl, Br, I, X$\neq$Y ) monolayers have been predicted with small exfoliation energy and excellent dynamical/thermal stabilities\cite{q16}. Here, we use GdBrI to construct $P$-symmetrical bilayer, as shown in \autoref{gd} (a) with interlayer  AFM ordering  being ground state.
To confirm our design principles, the layer-Gd-characteristics  projected  energy band structures of GdBrI for  monolayer,  $M_h$-symmetrical bilayer,  $P$-symmetrical bilayer  and   $P$-symmetrical bilayer  with $+z$ electric field  are plotted in \autoref{gd} (b, c, d, e) within SOC+$U$ ($U$=8 eV\cite{q16}). For monolayer GdBrI, the energy of the
K valley is higher than that of -K in the valence band maximum (VBM), resulting in a  valley splitting of 117 meV.
For $M_h$-symmetrical bilayer GdBrI, the energies of the K and -K valleys
are equal, suggesting  the disappeared valley polarization. However, the hidden valley polarization can be observed according to the layer characteristics of the enrgy bands. The hidden valley polarization can become observable by applying a perpendicular electric field (See FIG.S1\cite{bc}), which can induce an energy offset between the states of the upper and lower layers.
The hidden valley splitting of  each layer  is reduced to 113 meV. For $P$-symmetrical bilayer GdBrI, it is clearly seen that  the energy of the K valley is higher than that of -K  valley with valley splitting of 113 meV, giving rise to valley polarization.
For $P$-symmetrical bilayer  with $+z$ electric field, the degeneracy of spin or layer can be removed by  the layer-dependent electrostatic potential. When the direction of the electric field is reversed, the order of spin/layer splitting is also reversed (See FIG.S2\cite{bc}).

The $PT$ symmetry makes  each band be two-fold degenerate
with zero net Berry curvature in momentum space. However, each layer breaks the $PT$ symmetry individually, producing the layer-locked hidden Berry curvature\cite{f1}. Dut to spin-layer locking, the Berry curvatures between the spin-up and spin-down channels have the opposite value. That is, one spin channel has a positive value and the other possesses a negative value.
 The Berry curvature distributions of $P$-symmetrical bilayer GdBrI for the spin-up and spin-down channels are shown in \autoref{gd} (f, g). It is clearly seen that  the  berry curvatures are opposite  for different valley at the same spin channel and the same valley of different spin channel. With an applied longitudinal in-plane electric field,
the Bloch carriers of K valley by appropriate hole doping will acquire an anomalous transverse
velocity\cite{q4}, and the holes
are spontaneously deflected to opposite sides at different
layers  due to the layer-locked hidden Berry curvature, giving rise to valley layer-spin hall effect (\autoref{gd} (h)).
When a perpendicular
electric field  is applied, the system  can exhibit layer-locked AVHE   (\autoref{gd} (i, j)), which can be  tunable
by switching the electric field direction.

(2) $\mathrm{RuBr_2}$ monolayer: The  2H-$\mathrm{RuBr_2}$ monolayer has been predicted to be a ferrovalley material with excellent stability\cite{f2}. The $\mathrm{RuBr_2}$ for Monolayer,  $M_h$-symmetrical bilayer,  $P$-symmetrical bilayer  and   $P$-symmetrical bilayer  with  electric field shows similar results with Janus GdBrI. Here, we concentrate on the transition of valley polarization between valence and condition bands for $P$-symmetrical bilayer, along with its topological property. For monolayer $\mathrm{RuBr_2}$, it has been proven that both strain and electron correlation can induce transition of valley polarization between valence and condition bands, and can achieve valley-polarized quantum anomalous Hall insulator (VP-QAHI)\cite{f3}.
The $P$-symmetrical bilayer $\mathrm{RuBr_2}$ is shown in \autoref{ru} (a) with interlayer  AFM ordering  as ground state. Electron correlation effects on electronic structures of $P$-symmetrical bilayer $\mathrm{RuBr_2}$  are investigated by changing on-site Coulomb interaction $U$ value. Experimentally, the electronic correlation regulation can be achieved equivalently by strain engineering\cite{f3}.

The Ru-$d_{x^2-y^2}$/$d_{xy}$+$d_{z^2}$-orbitals projected  energy bands  with  $U=$1.5 eV, 2.0 eV and 2.5 eV are plotted in  \autoref{ru} (b, c, d). For all three $U$ values, $P$-symmetrical bilayer $\mathrm{RuBr_2}$  shows  valley polarization in both valence and condition bands. At $U$=1.5 eV, a remarkable valley splitting  can be observed in the  valence bands. However, the conduction bands show  the noteworthy valley polarization  at $U$=2.5 eV.  For $U$=2.0 eV, an intermediate phase appears.
 These can be explained  by the transformation  between $d_{z^2}$ and $d_{x^2-y^2}$/$d_{xy}$ orbitals.
The $d_{x^2-y^2}$/$d_{xy}$-orbital dominated  -K and K valleys have considerably larger  valley splitting than those with  $d_{z^2}$-orbital  characteristics\cite{q10,q11,q12,q13,q13-1,q14,q14-1,q15,q16,q18}.  At $U$=1.5 eV, the $d_{x^2-y^2}$/$d_{xy}$ ($d_{z^2}$) orbital dominates -K and K valleys of valence (conduction) bands.
For  $U$=2.5 eV, the distribution of Ru-$d$-orbitals is opposite to the  case of  $U$=1.5 eV.
In addition to  band inversion  between $d_{xy}$/$d_{x^2-y^2}$ and $d_{z^2}$ orbitals, the sign-reversible  Berry curvature can also be observed (See FIG.S3\cite{bc}) upon the occurrence of the transition of valley polarization  between valence and condition bands. These results of $P$-symmetrical bilayer $\mathrm{RuBr_2}$ are similar to those of monolayer case\cite{f3}.

\begin{figure}[t]
    \centering
    \includegraphics[width=0.45\textwidth]{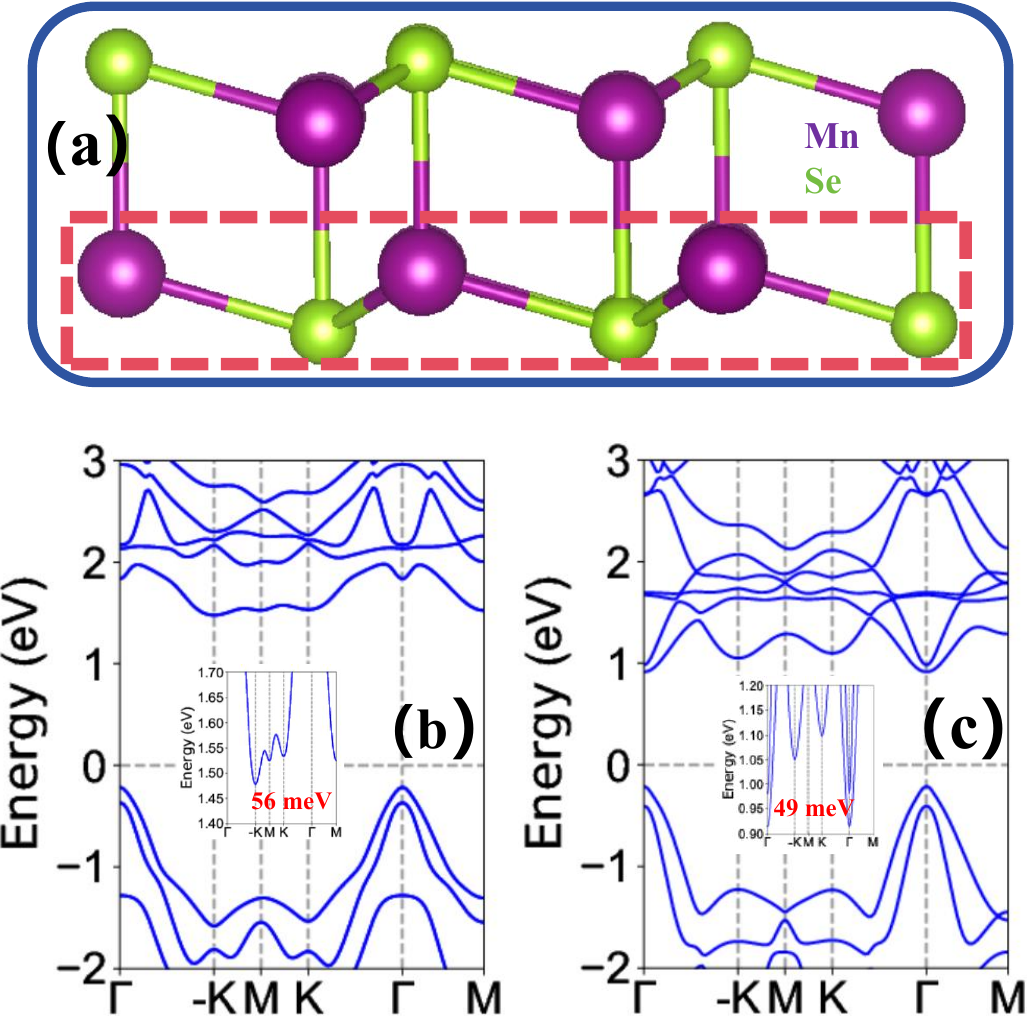}
     \caption{(Color online) For MnSe monolayer, (a): The crystal structures, being composed of two buckled honeycomb MnSe
sublayers, one of which is  marked by a red dotted box. (b, c): the energy band structures of MnSe monolayer and MnSe sublayer within SOC, and the insets show the enlarged portion of conduction bands near the Fermi energy level. }\label{mn}
\end{figure}
In a specific $U$ region, monolayer $\mathrm{RuBr_2}$  constitutes a quantum anomalous
Hall insulator (QAHI)\cite{f3}. A general design way for achieving 2D AFM QSHI has been proposed by stacking 2D  QAHI in a way that maintains the $PT$ symmetry\cite{f5}.  For bilayer $\mathrm{RuBr_2}$ with  $PT$ symmetry, the edge states  calculated  along the (100) direction with $U=$1.0 eV, 1.5 eV, 2.0 eV and 2.5 eV are plotted in \autoref{ru} (e, f, g, h).
It is found that, due to intrinsic spin mixing effects, a  \textit{near-ideal QSHI} can be realized with helical edge states and Dirac-like edge crossings with tiny gap for a specific $U$ region (for example: $U$=1.0 eV, 1.5 eV and 2.0 eV). We take $U$=1.5 eV as an example to elaborate, and its spin components of the edge states are plotted in \autoref{ru} (i). The $P$-symmetrical bilayer $\mathrm{RuBr_2}$  exhibits two pairs
of helical edge states with  Dirac-like edge crossing, and the second crossing only shows gap of about 3 meV.
For monolayer $\mathrm{RuBr_2}$,  it can be QAHI only in small areas around $U$=2 eV  (1.8 eV$<$$U$$<$ 2.105 eV)\cite{f3}.
The gap of Dirac-like edge crossing  depends on the strength of the SOC. When the $\mathrm{FeCl_2}$ monolayer is used as the basic building block to construct the $P$-symmetrical bilayer, the crossing gap is less than 1 meV (see FIG.S4\cite{bc}).

(3) Experimentally synthesized MnSe: The monolayer MnSe has been synthesized in
experiment, containing two FM MnSe
sublayers which are coupled antiferromagnetically\cite{f4}. As shown
in \autoref{mn} (a), the MnSe monolayer is composed of two buckled honeycomb MnSe
sublayers,  and they  are linked together by Mn-Se bonds.
The Mn/Se atoms of the upper sublayer are alternatively on top of  the Se/Mn atoms of the lower sublayer, leading to that MnSe
crystallizes in the  $P\bar{3}m1$ space group (No.~164),  hosting $P$ lattice symmetry.
If buckled honeycomb MnSe
sublayer is taken as the basic building block, monolayer MnSe satisfies our proposal in \autoref{sy} (c).

The valley polarization  generally exists in a $P$-broken hexagonal  monolayer.
MnSe has a $P$ lattice symmetry, and the $P$ symmetry will be broken only when the AFM ordering is considered.
In general, $P$-symmetry breaking induced by magnetic ordeing coupled with weak SOC leads to small valley splitting.
However, MnSe shows a relatively large valley spliting of 56 meV  in the conduction bands without strong SOC due to containing no heavy elements (see \autoref{mn} (b) with SOC+$U$ ($U$=2.3 eV\cite{f4})). This can be understood through our design principles, and the energy band structures of buckled honeycomb MnSe
sublayer is plotted in \autoref{mn} (c)). The MnSe
sublayer  exhibits a vally splitting of 49 meV, and a different band profile, especially in the conduction bands, compared with MnSe monolayer. These differences are because two  sublayers are linked together by  strong Mn-Se bonds, not weak vdW interaction.
Here, we explain the valley splitting of MnSe monolayer from a new perspective.

\textcolor[rgb]{0.00,0.00,1.00}{\textbf{Conclusion.---}}  In summary, we provide a  design way for bilayer with valley polarization from the AFM coupling between the two monolayers, based on ferrovalley monolayer and stacking operators.
When considering AFM ordering, the bottom and top layers of the bilayer are connected by $PT$-symmetry,  leading to valley polarization but spin/layer degeneracy. An out-of-plane electric field can be used to break $PT$ symmetry, thereby inducing spin/layer polarization.
Although the bilayer with an electric field becomes a fully compensated ferrimagnet, it can still be considered as an approximately ideal antiferromagnet, because a very small electric field can induce a considerable spin splitting due to large interlayer distance.
In fact, the fully compensated ferrimagnet inherits the advantage of the zero net magnetic moment of the antiferromagnet, leading to its intrinsic advantages of zero stray field and terahertz dynamics.
We demonstrate the effectiveness of our proposal  using Janus GdBrI monolayer as building block,  $\mathrm{RuBr_2}$ monolayer as building block, and  experimentally synthesized MnSe by the first-principles calculations.
A large amount of ferrovalley materials have been predicted, and the numerous candidate materials of AFM bilayer with valley polarization can be identified by utilizing the schemes presented in this work.
These results are beneficial for achiebing low-power and
high-efficiency valleytronics  devices with friendly paradigm.

\begin{acknowledgments}
This work is supported by Natural Science Basis Research Plan in Shaanxi Province of China  (2021JM-456). We are grateful to Shanxi Supercomputing Center of China, and the calculations were performed on TianHe-2.
\end{acknowledgments}

\end{document}